\documentclass[12pt,preprint]{aastex}
\usepackage{natbib}
\usepackage{subfigure}
\usepackage{times}
\begin{document}

\title{\emph{Fermi} LAT Observation of Supernova Remnant HB9}

\author{Miguel Araya}
\affil{Centro de Investigaciones Espaciales (CINESPA) \& Escuela de F\'isica,\\ Universidad de Costa Rica}
\affil{San Jos\'e 2060, Costa Rica}
\email{miguel.araya@ucr.ac.cr}

\slugcomment{Submitted to MNRAS}

\begin{abstract}
A 5.5-year \emph{Fermi} LAT gamma-ray observation shows significant extended emission at the position of the supernova remnant HB9 (G160.9+2.6). The significance of the detection above the background for photon energies above 0.2 GeV is 16$\sigma$. The gamma-ray flux above 0.2 GeV is $(2.23\pm{0.19_{stat}})\times 10^{-8}$ photons cm$^{-2}$s$^{-1}$, and the corresponding luminosity above 1 GeV is  $1.4 \times 10^{33}$ erg s$^{-1}$ (for a source distance of 1 kpc). The spectrum of the source is best described by curved power-law (log-parabola, $\frac{dN}{dE}=N_0 E^{-(\alpha+\beta \,\mbox{log}(E/1\,\mbox{\tiny GeV}))}$ with $\alpha = (2.24 \pm 0.09_{stat})$ and $\beta=0.4\pm 0.1_{stat})$). The gamma-ray spectrum of the source is consistent with both leptonic and hadronic models, and the relevant physical parameters in each case are derived. More studies on the ambient density in the region of HB9 should be carried out to rule out or confirm hadronic and non-thermal bremsstrahlung scenarios for the gamma-ray emission.
\end{abstract}


\section{Introduction}
Supernova remnants (SNRs) are believed to be the main sources of Galactic cosmic rays, charged particles with energies up to $10^{15}$ eV that are thought to gain their energy mainly through the mechanism of first order Fermi acceleration in the shocks of SNRs \citep{bel78,bla87} and its non-linear modification \cite[e.g.,][]{caprioli2012}. However, since these particles do not travel directly from their sources due to the presence of Galactic magnetic fields that change their directions of motion, the cosmic ray content of SNRs is studied indirectly by analyzing the properties of their non-thermal photon spectrum. High-energy electrons produce synchrotron emission that is seen mainly at radio wavelengths and up to X-ray energies \citep[see][for examples of X-ray spectra of non-thermal filaments in SNRs]{got01,ber02,hwa02,rho02,lon03,vink03,ber04}. These electrons may also produce leptonic gamma-ray photons through inverse Compton (IC) up-scattering of low energy ambient photons and bremsstrahlung emission \cite[e.g.,][]{gaisser1998}.

Gamma-rays may also result from hadronic interactions through the decay of neutral $\pi-$mesons produced in inelastic collisions between cosmic-ray protons (or ions) and ambient nuclei. The detection of a characteristic pion decay signature in the gamma-ray spectrum of SNRs would therefore constitute direct evidence for the existence of high-energy cosmic rays in these sources.

An important window of observations, with improved energy and angular resolution, mainly in the range from $\sim$ 100 MeV to 300 GeV, has been opened with the launch of the \emph{Fermi} observatory \citep{atw09}. However, most studies of gamma-ray observations of SNRs have yielded ambiguous results regarding the nature of the emission. SNRs that are relatively bright in gamma-rays and interact with molecular clouds, such as W28, W49B, W51C, G8.7-0.1 and, possibly, RCW 103, show spectra that seem to favour a hadronic origin \citep{abdo2010a,abdo2010b,abdo2009a,ajello2012,xing2014}. Recently, it has been claimed that, below $\sim$ 100 MeV, the spectrum of the SNRs IC 443 and W44 cannot be accounted for by leptonic mechanisms and shows a typical hadronic behaviour \citep{ackermann2013}. Other examples of GeV SNRs include the lower luminosity extended sources Cygnus Loop \citep{katagiri2011}, S147 \citep{katsuta2012} and G296.5+10.0 \citep{araya2013}; and younger SNRs, such as Cas A \citep{abdo2010c,araya2010}, Tycho's SNR \citep{giordano2012}, RCW 86 \citep{yuan2014}, RX J1713.7-3946 \citep{ellison2010,yuan2011} and RX J0852.0-4622 \citep{tanaka2011}.

In this paper, cumulative observations of the region around the SNR HB9 (G160.9+2.6) by the Large Area Telescope (LAT) onboard \emph{Fermi} are analyzed. HB9 is a large diameter (radio extension $120' \times 140'$), evolved, shell-type SNR showing thermal X-ray emission in the interior \citep{leahy1995} and fragmented radio shells \citep{leahy2007}. Based on X-ray observations, a distance of 1.5 kpc and an age of $8-20$ kyr are estimated \citep{leahy1995}, while a distance of $0.8\pm 0.4$ kpc is obtained from a likely association with HI structures \citep{leahy2007}. There are several other estimates of these parameters in the literature, for example, the evaporative cooling model developed by White \& Long \citep{white1991} give an age of 4-7 kyr and an explosion energy of $0.15-0.3 \times 10^{51}$ erg. An explosion energy of $0.2 \times 10^{51}$ erg, an age of 8 kyr and a distance of 1 kpc are adopted in this work.

The large extension of the SNR has allowed radio spectral studies in different regions of the source. After subtracting background radio sources, flux measurements have yielded a radio spectral index $\alpha = -0.51\pm 0.07$ ($S_\nu \propto \nu^\alpha$) between 408 and 1420 MHz \citep{kothes2006} and $\alpha=-0.47\pm 0.06$ \citep{leahy2007} for the remnant. Spatial variations in spectral index seen \citep{leahy2007} imply that the radio spectra of regions near the rim of HB9 are harder than those of regions towards the interior of the remnant, which is explained in terms of a higher compression and therefore higher magnetic field values at larger distances from the center of the SNR, which is also consistent with the higher radio brightness seen in the outer shell compared to the interior \citep{leahy2007}.

HB9 is believed to be located near the molecular clouds Sh217 and Sh219 \citep{leahy1991}, but no definitive interaction with a molecular cloud has been claimed. As has been mentioned, ambient gas interacting with SNRs might provide target material for cosmic-ray collisions, which is thought to be responsible for an enhancement of gamma-ray emission seen in some sources.

Several compact sources lie in the field of view of HB9 \cite[e.g.,][]{leahy1996}. The SNR might be associated to the pulsar PSR B0458+46 \cite[e.g.,][]{leahy1991}. Perhaps the most prominent source seen in the radio images of the region is the extragalactic radio source 0503+467 near the eastern edge of HB9 (source flux density of $\sim0.7$ Jy at 1.4 GHz, Spangler et al. 1987). This source has been observed with VLBI by Spangler et al. (1986, 1987) in search for predicted angular broadening by turbulence outside the shock of the SNR. The authors give a position for this source of RA (J1950)= 5$^h$03$^m$40$^s$.99, Dec (J1950)= +46$^\circ$41$'$46$''$.35 and its radio spectrum is reported (Spangler et al. 1987), which is seen to flatten due to synchrotron self-absorption at frequencies of 1-2 GHz.

The gamma-ray observation by the LAT of a region containing HB9 is described in Section \ref{observations}. The spatial and spectral properties of the observed emission are analyzed and its nature explored. The overall SED is modeled in Section \ref{model}. The discussion of results and conclusions are given in Section \ref{discussion}.

\section{\emph{Fermi} LAT Observation}\label{observations}
\emph{Fermi} LAT data between the beginning of the mission, 04 August 2008, and 28 February 2014 were analyzed with the standard software \emph{ScienceTools} version v9r32p5\footnote{See http://fermi.gsfc.nasa.gov/ssc} released October 24, 2013 with the latest reprocessed Pass7 photon and spacecraft data and the instrument response functions P7REP\_SOURCE\_V15. Standard selection criteria are applied to the data selecting \emph{Source} class events and a reconstructed zenith angle less than 100$^{\circ}$ to avoid contamination from gamma rays from Earth's limb. Recommended time intervals for data analysis are selected with the standard criteria (DATA\_QUAL==1) \&\& (LAT\_CONFIG==1) \&\& ABS(ROCK\_ANGLE)$<$52. The spectral analysis is restricted above 200 MeV due to uncertainties in the effective area and broad PSF at low energies, and below 100 GeV due to limited statistics.

Events within a square region of 14$^{\circ}\times$14$^{\circ}$ of the catalogued position of HB9, RA (J2000)= 5$^h$01$^m$00$^s$, Dec (J2000)= +46$^\circ$40, are used in the analysis to account for the large PSF. The emission model from this region includes the positions and spectral shapes of the point sources reported in the LAT 2-year Source Catalog \citep{nolan2012}. The recently released Galactic diffuse emission model (\emph{gll\_iem\_v05\_rev1.fit}) is used to account for the local extended background. This model corrects for anomalously low molecular hydrogen emissivities used in previous models, which improves the quality of the analysis mostly for Galactic sources in the longitude range $l=70^{\circ}$ to $l=300^{\circ}$. With the new Galactic background model, a considerable level of background emission that was previously unaccounted for was found to be particularly important near and at the position of HB9. The isotropic extragalactic file \emph{iso\_source\_v05.txt} is also used in the background model.

Spectral parameters of sources located beyond 10$^\circ$ of the position of HB9 are kept fixed to the values reported in the catalog. The fit is performed with the optimizer NEWMINUIT until convergence is achieved.

The spatial and spectral properties of LAT sources are obtained by means of a maximum likelihood analysis using the tool \emph{gtlike}\footnote{See http://fermi.gsfc.nasa.gov/ssc/data/analysis/scitools/binned\_likelihood\_tutorial.html}, which estimates the probability of obtaining the data given a source model and fits the parameters to maximize this probability. The previously detected \emph{Fermi} source 2FGL J0503.2+4643 which is seen towards the interior of the SNR is not included in the analysis. The resulting model is referred to as the null hypothesis. Another nearby gamma-ray source, 2FGL J0503.3+4517, lies at $1^{\circ}.4$ from the center of HB9, outside of the edge of the radio image. It has been associated with a background galaxy \citep{nolan2012}, and thus it is treated here as a background source.

The significance of a source is estimated by the test statistic (TS) defined as $-2$ log($L_0/L$), where $L_0$ and $L$ are the maximum likelihood values for the null hypothesis and for a model including the additional source, respectively \citep{mat96}. For large number of counts the detection significance of the source is given by $\sqrt{\mbox{TS}}$ \cite[however see][for some caveats]{protassov2002}.

\begin{figure}[ht]
\centering
\includegraphics[width=12.5cm,height=9cm]{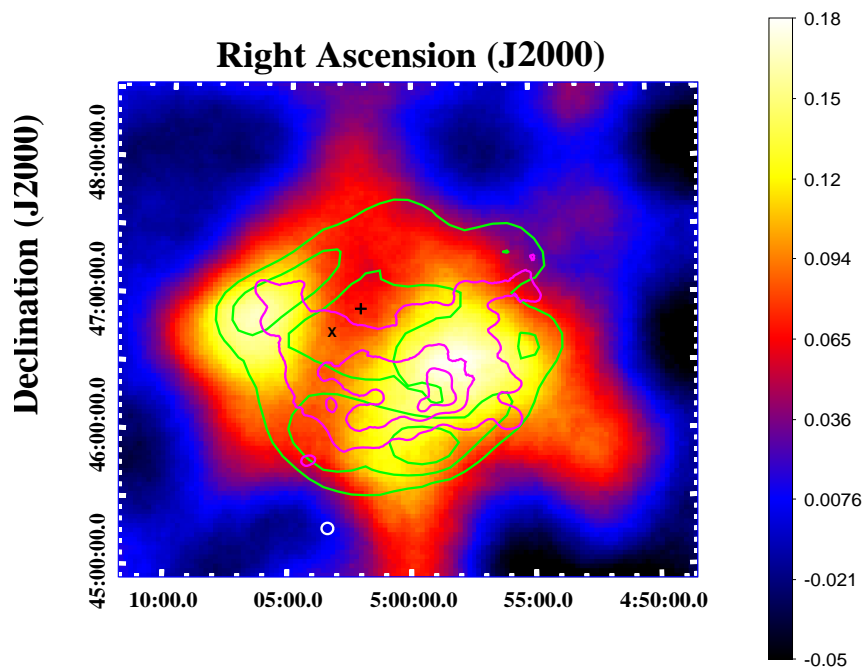}
\caption{Residuals map above 1 GeV, in units of counts per pixel, obtained after subtracting the background emission from the counts map, with a binning of 0.033 deg/pixel, smoothed using a Gaussian kernel with $\sigma=0^{\circ}.6$ (similar to the LAT PSF above 1 GeV). The `x' marks the position of the source 2FGL J0503.2+4643, which is not included in the model for HB9, the cross is the position of the pulsar PSR B0458+46 and the circle shows the position of the source 2FGL J0503.3+4517, associated with a background galaxy which is included in the model. The green contours are obtained from a GB6 4850 MHz image smoothed using a Gaussian kernel with $\sigma=0^{\circ}.3$; contours are at 0.6, 1.2, 1.8 and 2.4 kJy/sr. The magenta contours are obtained from a ROSAT image smoothed using a Gaussian kernel with $\sigma=0^{\circ}.1$; contour levels at 0.0001, 0.0002 and 0.0003 cts/s/pixel.\label{fig1}}
\end{figure}

In order to probe for new gamma-ray sources in the data, a `residuals map' is obtained above 1 GeV to take advantage of the narrower PSF at higher energies. The map is obtained by subtracting the observed counts map from the model map, which is constructed with the resulting likelihood fit to the data, and it is shown in Fig. \ref{fig1}.

Extended excess emission is apparent in Fig. \ref{fig1}, which also shows the radio and X-ray contours of the source obtained from a 4850 MHz Green Bank catalogue (GB6) image \citep{condon1994} and a ROSAT observation , respectively.

\subsection{Morphology and spectrum of the emission}\label{morphology}
Using events above 1 GeV, several hypotheses for the morphology of the residual gamma-ray emission are tested: spatial templates obtained from (a) the radio GB6 observation, (b) a ROSAT PSPC (X-ray) observation of HB9 and (c) a uniform disk template. The X-ray emission from HB9 is concentrated towards the center of the SNR \citep{leahy1995}, as shown in Fig. \ref{fig1}.

The center of the disk template, and its radius, are independently and systematically changed throughout the region of HB9 and the resulting TS value calculated. The highest TS for the disk is obtained for a template with a radius of $1^{\circ}.2$ $\pm 0^{\circ}.3$ and whose center is located within $0^{\circ}.3$ of the position RA (J2000)= 5$^h$01$^m$17$^s$, Dec (J2000)= +46$^\circ$24$'$49$''$. The quoted errors in the radius and position are at the 3$\sigma$ level.

The results of the morphological analysis are shown in Table \ref{table1}. The TS values with respect to the null hypothesis for the spatial morphologies above 1 GeV, assuming a power-law spectral shape above this energy, are (a) 91, (b) 64, and (c) 96.8 (in the same order as before). Although the uniform disk shows the highest TS value, the improvement to the fit is not significant compared to the following best-fit scenario (the radio template), since it has 3 additional degrees of freedom. Thus, the radio morphology is adopted here as a better description of the gamma-ray excess for the rest of the analysis.

The photon spectrum in the range 0.2-100 GeV is now obtained for the radio template. Different spectral shapes are explored and the results are shown in Table \ref{table2}. The spectral shapes probed are: a power-law ($dN/dE =N_0 E^{-\Gamma}$), a power-law with an exponential cutoff ($dN/dE= N_0 E^{-\Gamma}$exp$(-E/E_c)$) and a ``log-parabolic'' spectrum ($dN/dE = N_0 E^{-(\alpha + \beta \,\mbox{\small log}(E/1000\,\mbox{\small MeV}))}$). The resulting TS values are, respectively, 230, 255 and 261.

Evidence of spectral curvature is observed given the fit improvement using a log-parabola over a simple power-law. The highest TS value (261) corresponds to a spectral curvature significance of $\sim 5\sigma$ (see Table \ref{table2}). The fit parameters are $\alpha=(2.24 \pm 0.09)$ and $\beta=(0.4 \pm 0.1)$. The overall source significance for the best-fit spectral shape and spatial morphology is thus $16\sigma$, and its integrated photon flux and statistical error $(2.23\pm{0.19})\times 10^{-8}$ cm$^{-2}$s$^{-1}$ (0.2-100 GeV). Above 1 GeV, the corresponding gamma-ray luminosity is $1.4 \times 10^{33}$ erg s$^{-1}$ for a source distance of 1 kpc, about 40\% higher than the \emph{Fermi} LAT luminosity of the Cygnus Loop SNR above 1 GeV \cite{katagiri2011}, one of the dimmest gamma-ray SNRs detected so far, or about half the luminosity of G296.5+10.0 in the same energy range \citep{araya2013}. The model parameters can be seen in Table \ref{table2}.

The gamma-ray spectral energy distribution (SED) for the observed excess emission coincident with HB9 is obtained using the radio template to describe the morphology with a maximum likelihood analysis for 15 equally spaced logarithmic intervals from 0.2 to 100 GeV. If the significance of a detection within a bin is less than $3\sigma$ a 95\% confidence level upper limit for the flux is derived. Two sources of systematic errors are considered for each energy bin: the uncertainty of the Galactic diffuse emission which is estimated by varying the best-fit value of the normalization of the Galactic level by $\pm 6\%$ (Abdo et al. 2009a) and the uncertainty in the effective area, which is energy-dependent and given by $10\%$ at 100 MeV, $5\%$ at 560 MeV and $20\%$ at 10 GeV \citep{abdo2009b}. Fig. \ref{fig3} shows the SEDs with the data points obtained and statistical and systematic errors added in quadrature.

\subsection{Association}
Some pulsars are sources of gamma-ray emission \cite[e.g.,][]{abdo2013}. However, no point-like emission is seen at the position of the pulsar PSR B0458+46, as seen in Fig. \ref{fig1}, so this possibility for the origin of the emission is discarded here.

The extragalactic compact radio source 0503+467 lies at the boundary of the Eastern edge of the radio image of HB9. Radio galaxies are also known gamma-ray sources in the \emph{Fermi} band, and some of them show variability. A lightcurve was obtained here for a point source at the position of 0503+467 performing a binned likelihood analysis for events divided in time bins of $\sim 7$ months each (for such a time bin, the significance of the source is typically just above 3$\sigma$). No signs of flux variability in these time scales were found.

A more detailed analysis, which is beyond the scope of this study, should be carried out in order to confirm or discard the association of part of the gamma-ray emission seen in the region of HB9 to compact sources in the field. In what follows, an association to HB9 is assumed based on the likelihood results that show the emission is best described by an extended morphology.

\section{Emission Model}\label{model}
The radio SED is obtained from the literature \citep{reich2003,leahy2007,dwarakanath1982}. The radio spectrum can be described by synchrotron emission from a power-law energy ($\epsilon_e$) distribution of electrons varying as $\epsilon_e^{-2}$. The distribution extends up to a maximum electron energy, $\epsilon_{e,m}$, and the particles are uniformly distributed across the volume of the radio shell of the SNR (which is approximated as a sphere of radius 18 pc). The measured X-ray (0.1-2.5 keV) flux from HB9 \citep{leahy1995}, emitted by the hot gas inside the shell, is used as an upper limit for the synchrotron emission in this energy band.

Gamma-ray emission mechanisms include IC up-scattering of Cosmic Microwave Background (CMB) photons, bremsstrahlung emission from the synchrotron-emitting electrons interacting with ambient plasma \cite[assumed to consist only of thermal electrons, protons and ionized helium with standard abundances and proton density $n_p$, see][]{araya2010} and neutral pion decay produced in interactions of cosmic ray protons with ambient protons (with density $n_p$). Any contribution to IC emission from other photon fields such as infrared emission from dust, or interstellar radiation fields, is not included in this model.

Hadronic gamma-ray emission is calculated according to Aharonian \& Atoyan \citep{aharonian2000} for a particle density distribution in the form of a power-law in \emph{momentum} ($p$) of the form $p^{-s}$, as predicted by the theory of diffusive shock acceleration. Note that a pure power-law in momentum results in a slightly ``curved'' particle flux in kinetic energy ($T$) of the form $\beta \,(dp/dT)\, p^{-s}\,\propto (T^2 + 2mc^2T)^{-s/2}$ \cite{dermer2012}, which is more significant at lower particle energies but should be taken into account in the calculations if one works with a distribution in kinetic energies.

\subsection{IC-dominated scenario}
The gamma-ray emission can be explained with a simple power-law electron energy distribution, as can be seen in Fig. \ref{fig3}. The electron distribution extends up to $\epsilon_{max,e}= 0.5$ TeV. The magnetic field and total electron energy are given, respectively, by 8 $\mu$G and $5.4\times 10^{48}$ erg, or 2.7\% of the total explosion energy of $2\times 10^{50}$ erg.

Previous X-ray observations \citep{leahy1995} imply low ambient densities. The model shown in Fig. \ref{fig3} for this scenario shows the corresponding bremsstrahlung flux with an ambient density $n_p=0.1$ cm$^{-3}$.

\subsection{Hadronic-dominated scenario}
Since the gamma-ray spectrum steepens above $\sim$1-2 GeV, a proton spectrum with a power-law momentum distribution with index $s=2.6$ is able to fit the SED. However, such a particle distribution is rather steep, and a more physically plausible scenario is obtained with a particle distribution of the form $p^{-2.3}$ exp[$(-p/80\,\mbox{GeV/c})$], which is also shown in Fig. \ref{fig3}. The break in the particle momentum distribution is required by the observations. A lower index below the break of 2, like that of electrons, results in gamma-ray emission that is too hard to account for the \emph{Fermi} LAT fluxes.

The total cosmic ray energy, $E_{CR}$, and ambient density in this case are given by $E_{CR}=4\times 10^{49}\left(\frac{n_p}{1\,\mbox{\tiny cm}^{-3}}\right)^{-1}$ erg. If, for example, $\sim$20\% of the available energy is in these cosmic rays, the required density is $n_p=$1 cm$^{-3}$. This scenario is shown in Fig. \ref{fig3}, where the leptonic component is calculated with a magnetic field of 40 $\mu$G, a power-law electron distribution with index 2 extending up to 460 GeV and a total electron energy of $5\times 10^{47}$ erg. An ambient density of 1 cm$^{-3}$ is used for the bremsstrahlung flux also.

\subsection{Bremsstrahlung-dominated scenario}
The gamma-ray flux upper limits constrain the maximum electron energy in this scenario, which should not exceed 12 GeV. Together with an ambient density of 1 cm$^{-3}$, a magnetic field of 6 $\mu$G and total electron energy content of $6.7\times 10^{48}$ erg, the modeled SED is also shown in Fig. \ref{fig3}.

\section{Discussion and Conclusions}\label{discussion}
Excess gamma-ray emission in the region of the large SNR HB9 is revealed by analysis of cumulative observations by \emph{Fermi} LAT. The morphology of the gamma-rays is best described by a the radio morphology of the source (see Table \ref{table1}). In the energy band explored (above 0.2 GeV) the photon spectrum is curved and can be best described by a log-parabola of the form $E^{-(2.24+ 0.4 \,\mbox{log}(E/1\,\mbox{\tiny GeV}))}$.

The extension of the SNR offers an exciting opportunity to carry out studies on spatially-dependent particle acceleration. For example, a spatial study of the radio emission in the band 408-1420 MHz has shown that synchrotron emission from the center of the shell is softer than that from the outer shell \citep{leahy2007}. However, the current statistics of the gamma-ray observation is too low to draw definitive conclusions regarding any possible variation of the flux level or of the spectral parameters across the source.

\begin{figure}[ht]
\centering
\includegraphics[width=10cm,height=5.5cm]{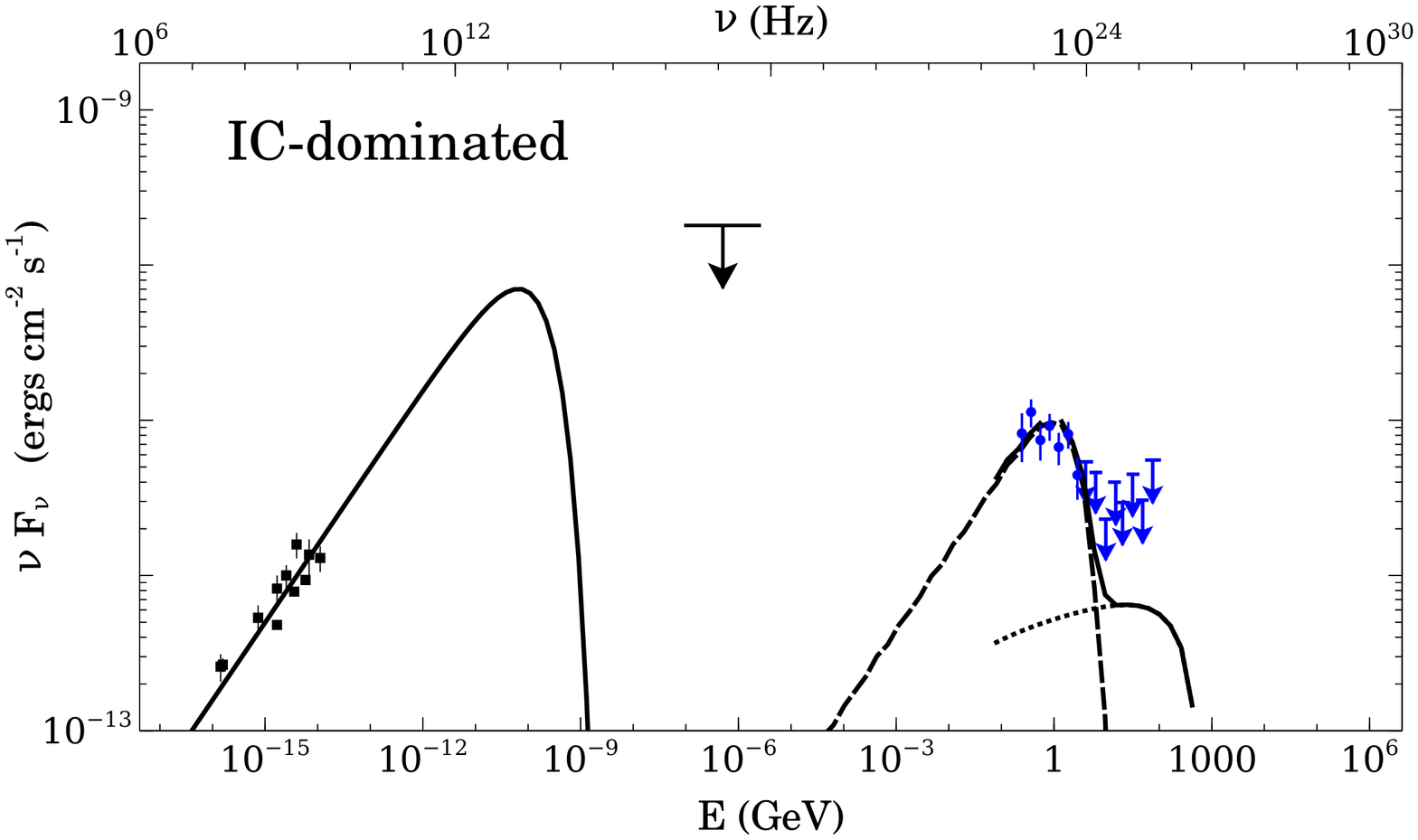}
\includegraphics[width=10cm,height=5.5cm]{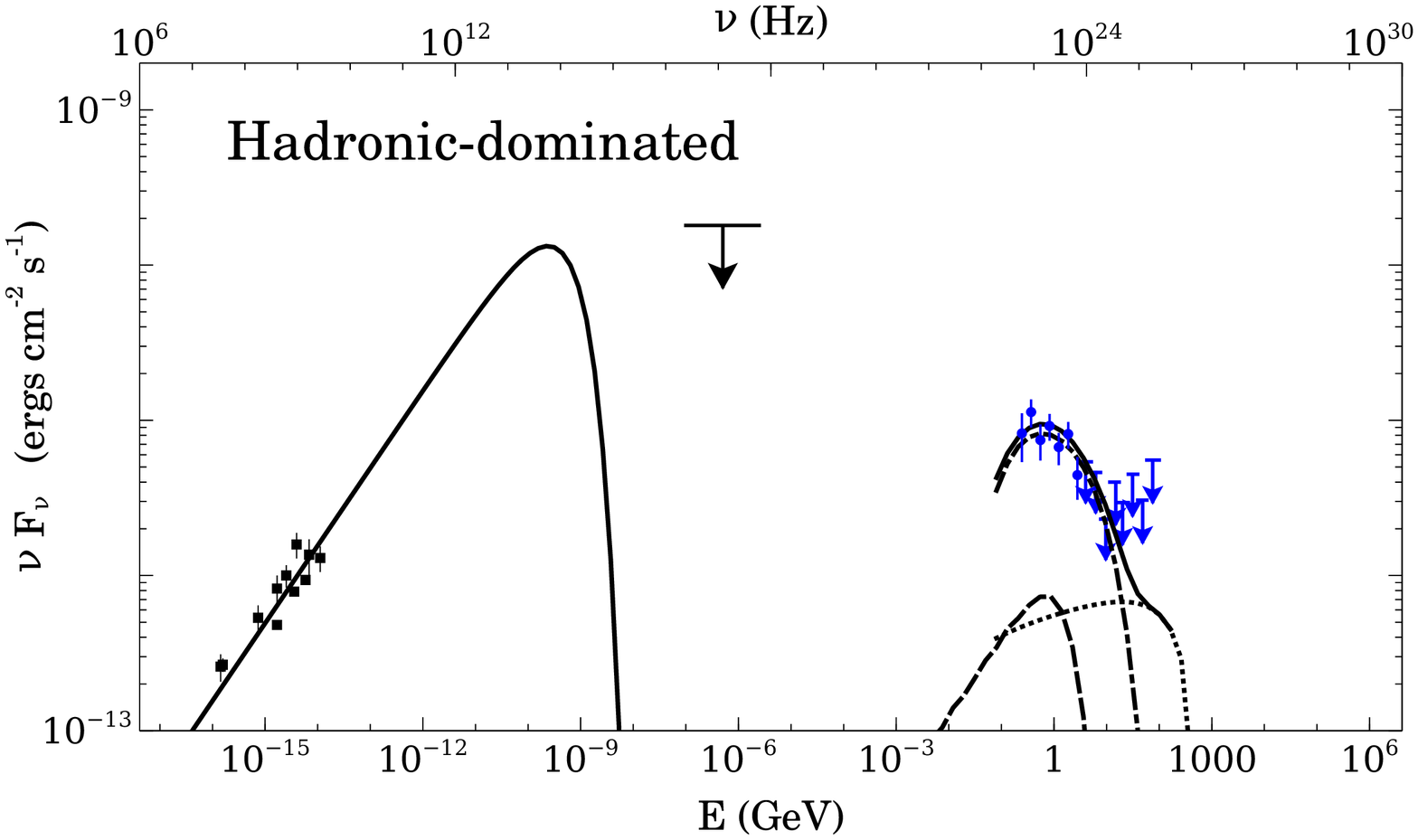}
\includegraphics[width=10cm,height=5.5cm]{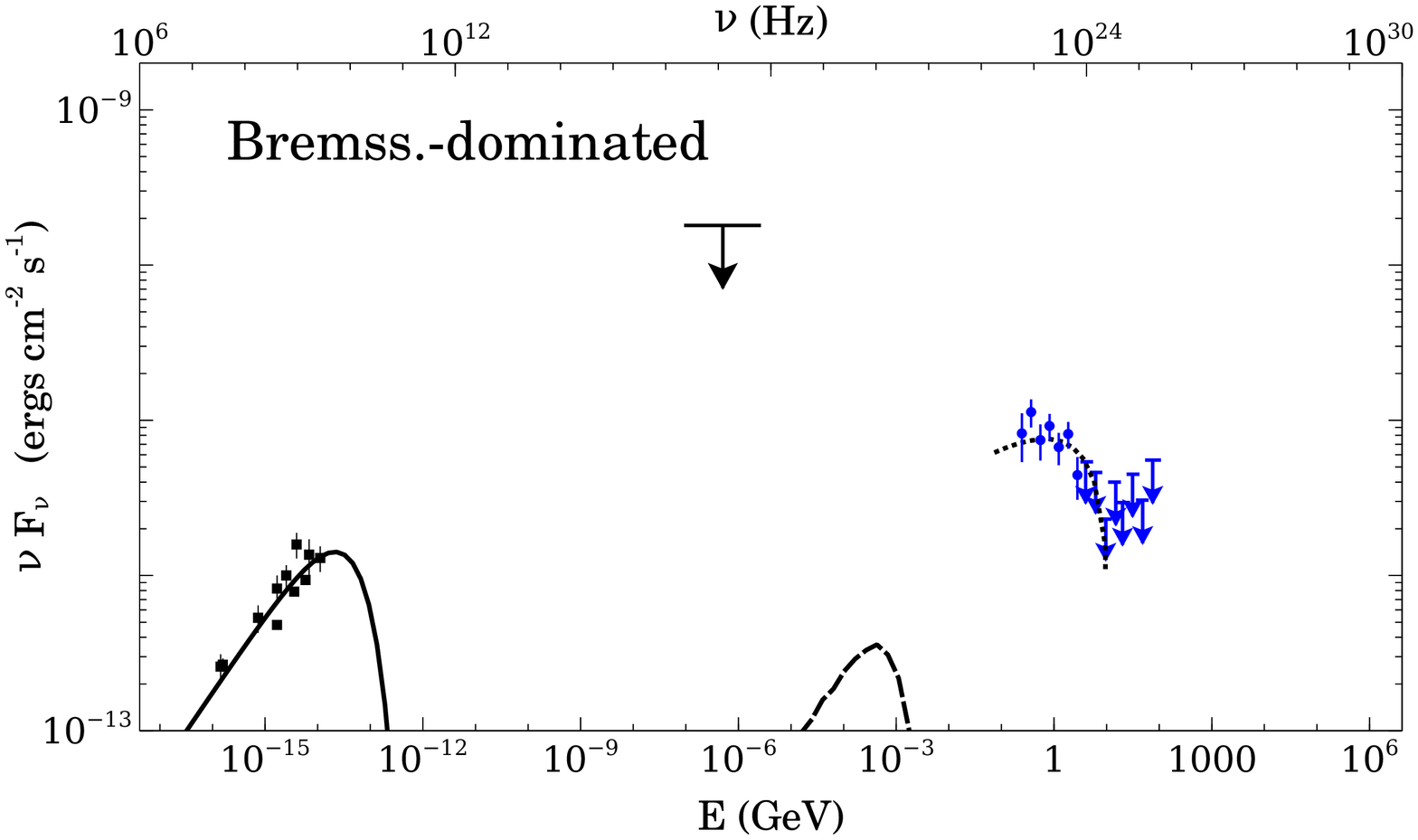}
\caption{Emission scenarios for the broadband SED of HB9. The components are: synchrotron (solid line), IC-CMB (dashed line), non-thermal bremsstrahlung (dotted line), neutral pion decay (dash-dotted line) and total gamma-ray emission (solid line in the IC and hadronic scenarios). Blue circles are obtained from the LAT observation in this paper. Upper limits at the 95\% confidence level are shown for intervals with no source detection.}
\label{fig3}
\end{figure}

Different mechanisms for the origin of the gamma-ray emission from HB9 have been explored, and the corresponding models are shown in Fig. \ref{fig3}.

The gamma-ray spectrum is consistent with an inverse Compton scenario where CMB photons are up-scattered by the synchrotron-emitting electrons responsible for the radio emission. The physical parameters derived are reasonable (magnetic field 8 $\mu$G, total electron energy of 2.7\% of the available energy) and consistent with observations in the X-ray band (low ambient density, $\sim 0.05$ cm$^{-3}$). The gamma-ray data observed above 0.2 GeV corresponds, in this scenario, to the high-energy tail of the IC emission. The electron distribution that is required to account for the LAT
observation is a simple power-law in particle energy with an index of 2 and maximum electron energy of 0.5 TeV. The maximum energy in the electron spectrum contains information on the details of the acceleration process. In the absence of escape, the maximum energy can be determined by radiative losses after a sufficiently long time. Synchrotron losses will dominate over IC-CMB losses for magnetic fields above $\sim$3.27 $\mu$G, for which the magnetic energy density is the same as the CMB energy density \citep{reynolds1998}. The maximum electron energy in this case is then given by \citep{reynolds1998} $\epsilon_{e,m}\approx (30\,\,\mbox{GeV})\,u_3/\sqrt{\eta R_J B_1}$, where $B_1$ is the upstream magnetic field in Gauss, $u_3$ is the shock speed in units of $10^3$ km/s, $\eta$ is the ratio of the particle mean free path to its gyroradius (also a measure of the amount of magnetic turbulence), and $R_J$ is a factor depending on the shock obliquity, the shock compression ratio and $\eta$.

For a SNR age of 8 kyr, explosion energy of $0.2\times 10^{51}$ erg and ISM density of 0.1 cm$^{-3}$, the Sedov-Taylor solution gives a shock speed $u_3 =0.74$. If the ISM magnetic field is $B_1=3\times 10^{-6}$ G then, for a parallel shock with $R_J=1$, the maximum electron energy found in this work through a fit of the gamma-ray SED in the IC-dominated scenario gives $\eta \approx 660$, which would imply very low resonant magnetic turbulence ($\delta B/B=1/\sqrt{\eta}\sim0.04$).

A hadronic interpretation might seem less likely due to the low ambient densities observed. The required mean density across the SNR is around 1 cm$^{-3}$ for a total cosmic ray energy of 20\% of the available energy of $0.2\times10^{51}$ erg (or higher density for a corresponding lower cosmic ray energy content). A power-law with exponential cutoff momentum distribution with index $s=2.3$ and cutoff momentum 80 GeV/c for the protons can account for the gamma-ray SED.

In the bremsstrahlung scenario, a similar value for the density is used (1 cm$^{-3}$). The electron distribution in this case must extend up to a particle energy of only 12 GeV, which is close to the lower limit for the maximum electron energy required to explain the radio data. Extending the radio observations of this object above 3 GHz could discard this scenario if the same power-law synchrotron emission is present at higher radio frequencies. The high density requirement can be relaxed by increasing the total energy in the electrons and lowering the magnetic field (the value used is 6 $\mu$G). However, the magnetic field is likely not lower than at least the typical Galactic value ($\sim$3 $\mu$G).

For both the hadronic and bremsstrahlung scenarios the density requirements could be easily achieved if the SNR interacts with clumps of high density material in its environment. New observations should be carried out to confirm or discard this possibility.

Finally, it is noted that a better fit to the SED might be obtained with more complex scenarios such as those involving several particle populations. Mixed scenarios involving comparable fluxes from leptonic and hadronic emission mechanisms are also certainly plausible and have not been explored here.

\acknowledgments
This research has made use of NASA's Astrophysical Data System and of the SIMBAD database. Financial support from Universidad de Costa Rica, through research grant number 829-B2-171 from Vicerrector\'ia de Investigaci\'on, and from the Escuela de F\'isica, is acknowledged. Comments made by the anonymous referee were highly appreciated and helped improved the quality of this work substantially.

\begin{table}
\begin{center}
\caption{Test Statistic for Spatial Models Compared to the Null Hypothesis for HB9 (1-100 GeV)\label{table1}}
\begin{tabular}{|c|c|c|}
\tableline\tableline
 & Test Statistic & Additional Degrees of Freedom\\
\tableline
Null hypothesis (background only) & 0 & 0\\
4850 MHz & 91 & 2\\
ROSAT (X-ray)& 64 &2\\
Uniform disk$^{a}$ & 96.8 & 5\\
\tableline
\end{tabular}
\tablenotetext{a} {Best fit radius of 1$^{\circ}$.2$\pm 0^{\circ}.3$ and center at RA (J2000)= 5$^h$01$^m$17$^s$, Dec (J2000)= +46$^\circ$24$'$49$''$.}
\end{center}
\end{table}

\begin{table}
\begin{center}
\caption{Best-Fit Photon Spectral Parameters for the Radio Template (0.2-100 GeV)\label{table2}}
\begin{tabular}{|c|c|c|c|}
\tableline\tableline
Spectral shape & Photon flux (cm$^{-2}$s$^{-1}$)& Parameter values & TS\\
\tableline
$\frac{dN}{dE}=N_0 E^{-\Gamma}$ & $(2.42\pm 0.19)\times 10^{-8}$ & $\Gamma=2.30\pm 0.05$& 230\\
& & &\\
\tableline
$\frac{dN}{dE}=N_0 E^{-\Gamma}\mbox{exp}\left(-\frac{E}{E_c}\right)$& $(2.24\pm 0.20)\times 10^{-8}$ & $\Gamma=1.44\pm{0.25}$& 255\\
&  & $E_c=(1.6\pm 0.6)$ \small{GeV} & \\
& & &\\
\tableline
$\frac{dN}{dE}=N_0 E^{-(\alpha+\beta \,\mbox{log}(E/1000\,\mbox{\tiny MeV}))}$ & $(2.23\pm{0.19})\times 10^{-8}$ &$\alpha=2.24\pm0.09$ & 261\\
& &$\beta=0.4\pm0.1$ & \\
\tableline
\end{tabular}
\tablenotetext{a} {Only statistical errors are shown.}
\end{center}
\end{table}

\end{document}